\documentclass[onecolumn,english,10pt, conference, compsocconf]{IEEEtran}
\usepackage[T1]{fontenc}
\usepackage[latin9]{inputenc}
\usepackage{colortbl}
\usepackage{array}
\usepackage{float}
\usepackage{multirow}
\usepackage{amsmath}
\usepackage{amssymb}
\usepackage{graphicx}

\makeatletter
 
\providecommand{\tabularnewline}{\\}
\floatstyle{ruled}
\newfloat{algorithm}{tbp}{loa}
\providecommand{\algorithmname}{Algorithm}
\floatname{algorithm}{\protect\algorithmname}

\ifCLASSINFOpdf
\else
\fi
\makeatother

\usepackage{babel}
\begin{document}
\title{FPGA-based Matched Filter Group Optimisation for SKA Pulsar Search
Engine}
\author{\IEEEauthorblockN{Haomiao Wang\IEEEauthorrefmark{1}, Ben Stappers\IEEEauthorrefmark{2}, Prabu Thiagaraj\IEEEauthorrefmark{3},  
and Oliver Sinnen\IEEEauthorrefmark{1}}\\
\IEEEauthorblockA{\IEEEauthorrefmark{1}PARC Lab, University of Auckland\\
\IEEEauthorrefmark{2}Jodrell Bank Centre for Astrophysics, University of Manchester\\
\IEEEauthorrefmark{3} Raman Research Institute, University of Manchester\\ 
hwan938@aucklanduni.ac.nz, ben.stappers@manchester.ac.uk\\ prabuthiagaraj@gmail.com, o.sinnen@auckland.ac.nz} }
\maketitle
\begin{abstract}
Pulsar search is one of the main tasks for the Square Kilometre Array~(SKA),
implemented in the central signal processor~(CSP) sub-element. As
most the characteristics of undiscovered pulsars are unknown by definition,
exhaustive searches over a multi-dimensional parameter space are employed.
One main compute-intensive task of the pulsar search modules in the
CPS is the matched filter group, which convolves the input signals
with a group of large FIR filters. High-performance designs on FPGAs
have been proposed that can process multiple large filters efficiently.
But given that in many applications, including the here targeted pulsar
search, FIR filters have many different sizes, there is further potential
for optimisation. This paper investigates the optimisation of matched
filtering designs. While the results are tranferable to other domains,
we are motivated by the needs of the SKA pulsar search engine. The
influence of changing number of filters and the difference in sizes
is analysed. The generic design in time-domain~(TD) is optimised
by employing the longest processing time~(LPT) first rule to distribute
filter templates across filter processing pipelines. For the Fourier-domain~(FD),
the relationship between the required off-chip memory space and speedup
over the generic design is investigated. To put the results into relation
with with GPU design, we compared with a well-optimised design for
top-end GPUs (NVIDIA Tesal P100). While a mid-range Intel Arria 10
is up to 7.5x slower than the P100, the performance per watt is slightly
better on the Arria 10.
\end{abstract}

\section{Introduction}

The scale of the Square Kilometre Array~(SKA)~\footnote{www.skatelescope.org}
is tens of times larger than existing radio telescope arrays, and
so is the number of signals that will be received and processed by
the SKA~\cite{dewdney2009square}. This is a big challenge for the
SKA central signal processor~(CSP). One of the main elements in the
CSP is the pulsar search engine~(PSS) that employs several approaches
to search for different types of pulsars. Since the characteristic sof
undiscovered pulsars are unknown by definition, the PSS has to brute
force search a wide range of values for each relevant parameter. 

The matched filtering technique is an important and efficient signal
processing approach to recover the specific polluted signals from
raw received signals~\cite{ransom2002fourier}. Apart from other
signal processing areas, it is widely employed in the pulsar search
modules. For the SKA1-MID pulsar search engine~\cite{ransom2011presto},
it appears in many modules such as single pulsar detection~(SPDT),
Fourier domain acceleration search~(FDAS), and folding and optimisation~(FLDO).
However, the employed matched filter groups are not the same, which
is also true for the number of filters and the incremental of the
number of filter coefficients/taps. 

An important feature of the SKA pulsar search module is that the input
data size is hundreds of times larger than the output data size, and
the input data cannot be stored in FPGA on-chip memory. For example,
the input signals sizes to SPDT, TDAS, and FDAS are millions of points.
However, the output signals are all a list of pulsar candidates, whose
sizes are less than 1\% of those of input signals. Because of the
huge amount of workload and restricting time limitation (new signals/data
is constantly coming in), the needed throughput is high, and high-end
acceleration devices are required. This makes high-end FPGAs interesting
and efficient candidates for hardware acceleration of matched filter
group processing. 

In this research, we investigate the FPGA-based optimisation of matched
filtering designs with different sizes. The main contributions are
as follows:
\begin{itemize}
\item Analysing and modelling of of matched filter groups of different sizes
in time-domain and Fourier-domain; 
\item Optimising the general FPGA-based time-domain and Fourier-domain matched
filter group designs using load balance algorithms;
\item Providing the best domain and parameters for any matched filter group
based on the execution latency and logic resources on a specific device;
\item Evaluation of the best design space parameters for time-domain designs
and Fourier-domain designs; evaluation of the benefit of optimised
matched filter groups designs in comparison to commonly used generic
implementations; comparison of the best performing FPGA design with
well-optimised design on high-end GPU. 
\end{itemize}
The rest of the paper is organized as follows. Section~\ref{sec:Related_work}
gives background on matched filtering in radio astronomy and FPGA
as an accelerator. Section~\ref{sec:Matched_Filtering} discusses
matched filtering, the design goals, and gives a theoretical analysis.
In Section~\ref{sec:Optimistion}, the proposed designs for matched
filter groups are discussed, and the load balance algorithm is employed
for multiple processing units. Section~\ref{sec:Evaluation} presents
the evaluation and discusses the results. Finally, the conclusions
are given in Section~\ref{sec:Conclusions}. 

\section{\label{sec:Related_work}Related Work}

\subsection{FPGA as an Accelerator}

Most of the real-time compute-intensive applications in radio astronomy
projects cannot be handled by general high-end processors, and acceleration
devices are widely employed~\cite{parsons2009digital}. High-end
GPUs are strong competitors tohigh-end FPGAs in hardware acceleration~\cite{sanchez2005digital}.
Because FPGAs have advantages over GPUs in terms of power efficiency,
i.e. in performance per watt, they are adopted in many digital signal
processing modules of science projects. For the high-speed data-filtering
in the CERN project~\cite{sridharan2016accelerating}, an FPGA is
3.4x times better than a high-end GPU (Titan X Pascal) and 2.3x times
better in terms of performance/watt. Hundreds of Xilinx Virtex-4 FPGAs
are installed to accelerate the correlator of the SKAMP project~\cite{de2007radio}.
An FPGA-based accelerator appears in the FX correlator of the radio
telescope MeerKAT as well~\cite{parsons2009digital}. In~\cite{sanchez2005digital},
high-end FPGA platforms are employed to handle digital channelised
receivers. 

\subsection{Convolution Acceleration}

High-end FPGAs are widely employed as accelerators in the machine
learning area~\cite{nakahara2017object} such as the convolutional
neural networks (CNN)~\cite{zeng2018framework} and deep learning~\cite{aydonat2017opencl}.
The main compute-intensive task for these applicationsis convolution.
Although some accelerations are for 2D convolution, FPGAs also provide
good performance for accelerating 1D convolution. In~\cite{fowers2013performance},
this is evaluated on different hardware platforms, and FPGA devices
perform better than CPUs and GPUs when there are several hundred filter
coefficients (or taps). In~\cite{wang2016fpga}, efficient FIR filter
designs for FPGAsare investigated in both time-domain and Fourier-domain,
and multiple FPGA devices can provide better performance in terms
of execution latency, while consuming less energy than a GPU. Regarding
the matched filter technique in pulsar search, it is widely employed
and implemented using FPGAs~\cite{ransom2002fourier}.

\section{\label{sec:Matched_Filtering}Matched Filtering }

A matched filter group is employed to correlate a group of known templates
with unknown signals to detect the presence of a template in the signals~\cite{turin1960introduction}\@.
For pulsar search, the characteristics of an undiscovered pulsar are
unknown, and a group of predicted templates are employed. In essence,
the signal array is convolved with $N_{filter}$ templates, which
can be presented as

\begin{align}
y[i][j] & =\sum^{Tap_{i}}_{k=1}x[j-k]h[i][k],\label{eq:tdfir}\\
for & \,i=1,2,\thinspace...N_{filter\,}and\,j=1,2,\,...N_{input}\nonumber 
\end{align}
where $y$ is the filter output plane, $x$ is the input array, $h$
are the coefficient arrays, and $Tap_{i}$ is the length of the $i$th
filter (coefficients are sometimes called taps and both words are
used interchangeably in the following). Regarding the input and coefficient
arrays in the SKA project, all the data types are complex single-precision
floating-point (SF) numbers. The lengths from filter$_{1}$ to filter$_{N_{filter}}$
varies, and the filter lengths are given by
\[
Tap_{i+1}=Tap_{i}+N_{inc}
\]
\[
Tap_{i}=Tap_{1}+N_{inc}(i-1),
\]
 where $N_{inc}$ is tap incremental, i.e. the difference in size
of two consecutive filters. There are three factors that characterise
a matched filter group: 1) the number of filters~$N_{filter}$, 2)~the
tap incremental $N_{inc}$, and 3)~range of filter lengths {[}$Tap_{1},\,Tap_{N_{filter}}${]}.
In this research, we use $MF-(N_{filter},\,N_{inc},\,Tap_{1})$ to
represent a matched filter group; $Tap_{N_{filter}}$ can be calculated
from the other values.

Table~\ref{tab:Matched-filter-group} summarises the parameters of
a matched filter group and lists the general range of each parameter
considered in this work.

\begin{table*}
\centering{}\caption{\label{tab:Matched-filter-group}Matched filter group parameters}
\begin{tabular}{|c|c|c|}
\hline 
Parameter & Description & Range\tabularnewline
\hline 
\hline 
$N_{filter}$ & Number of FIR filters & $[1,\,1000]$\tabularnewline
\hline 
$Tap_{i}$ & Length of the $i$th FIR filter & $[1,\,1000]$\tabularnewline
\hline 
$N_{inc}$ & Value of tap incremental & $[1,\,10]$\tabularnewline
\hline 
$N_{input}$ & Length of input array & $[2^{10},\,2^{30}]$\tabularnewline
\hline 
\end{tabular}
\end{table*}

\subsection{\label{subsec:Matched-Filtering}FIR Filter and Algorithms}

The matched filter group can be seen as a group of FIR filters. Each
filter of the group can be implemented in both time-domain (TDFIR),
Equation~\ref{eq:tdfir}, and Fourier-domain (FDFIR), Equation~\ref{eq:fdfir}.
\begin{equation}
x\ast h=\mathcal{F}^{-1}\{\mathcal{F}\{x\}\cdot\mathcal{F}\{h\}\},\label{eq:fdfir}
\end{equation}
where $\mathcal{F}\{\cdot\}$ and $\mathcal{F}^{-1}\{\cdot\}$ are
Fourier transform and inverse Fourier transform~\cite{smith1997scientist},
respectively. 

\begin{figure*}
\begin{centering}
\includegraphics[viewport=25bp 20bp 430bp 220bp,clip,scale=0.6]{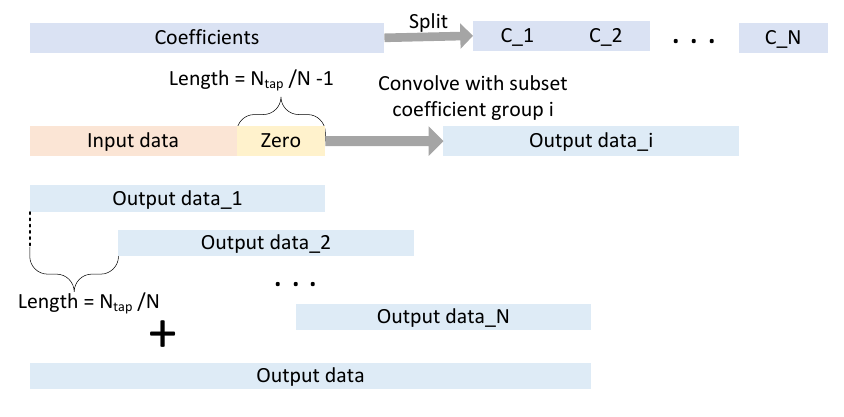}\includegraphics[viewport=25bp 20bp 400bp 290bp,clip,scale=0.6]{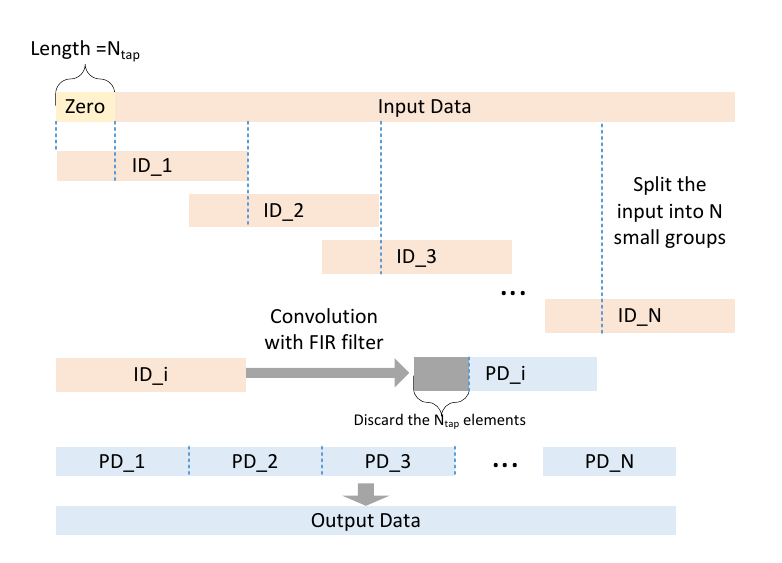}
\par\end{centering}
\centering{}\caption{\label{fig:ola-ols}Process flow of the overlap-add algorithm (top)
and the overlap-save algorithm (bottom)}
\end{figure*}

For the large tap TDFIR, the overlap-add (OLA) algorithm~\cite{smith1997scientist},
illustrated in Figure~\ref{fig:ola-ols}(top), can be employed to
split the FIR taps into a group of small filters, each of size $N_{OLA-tap}$
, when not all taps can processed simultaneously, e.g. due to the
lack of resources in an FPGA. Correspondingly for the FDFIR filter,
the overlap-save (OLS) algorithm~\cite{smith1997scientist}, illustrated
in Figure~\ref{fig:ola-ols}(bottom), can be employed to divide the
input array into a set of small input arrays~\cite{pavel2013algorithms}
when the number of inputs, i.e. FFT points, is too large, such as
several millions.

\subsection{Design Goals}

For FPGA-based acceleration, the initial FPGA configuration time can
be over one second, and even the partial reconfiguration takes tens
to a hundred of $ms$. To avoid reconfiguration, all filters of the
group are implemented using a single FPGA image (bitstream file).
In a generic implementation~\cite{wang2016fpga,wang2019fpga}, the
number of filter taps of all filters are the same, which is $N_{tap_{N_{filter}}}$,
and one design can implement all filters. If a filter is smaller,
i.e. has less taps, than $N_{tap_{N_{filter}}}$, it is padded with
zeros to make its length to $N_{tap_{N_{filter}}}$. Since many filters
are padded, a large proportion of operations are unnecessary. In this
research, we investigate the optimisation of the generic FPGA-based
matched filtering implementation. The main goal is to propose a generalised
design for any given matched filter group targetting a specific FPGA. 

\subsection{\label{subsec:Theoretical-Analysis}Theoretical Analysis}

\subsubsection{Time-Domain~(TD) Filtering}

Based on Section~\ref{sec:Matched_Filtering}, the total number of
taps of an entire filter group is 
\[
\sum^{N_{filter}}_{i=1}Tap_{i}=N_{filter}Tap_{1}+\frac{N_{inc}N_{filter}(N_{filter}-1)}{2}.
\]
 For a generic matched filter group implementation, the filter lengths
are padded to be the same as $Tap_{N_{filter}}$, and the sum of all
taps is 
\[
Tap_{N_{filter}}N_{filter}=N_{filter}Tap_{1}+N_{inc}N_{filter}(N_{filter}-1).
\]
The difference of these two values is $N_{inc}N_{filter}(N_{filter}-1)/2$,
which is up to half of $Tap_{N_{filter}}N_{filter}$. 

\begin{figure*}
\begin{centering}
\includegraphics[viewport=0bp 218bp 420bp 418bp,clip,scale=0.6]{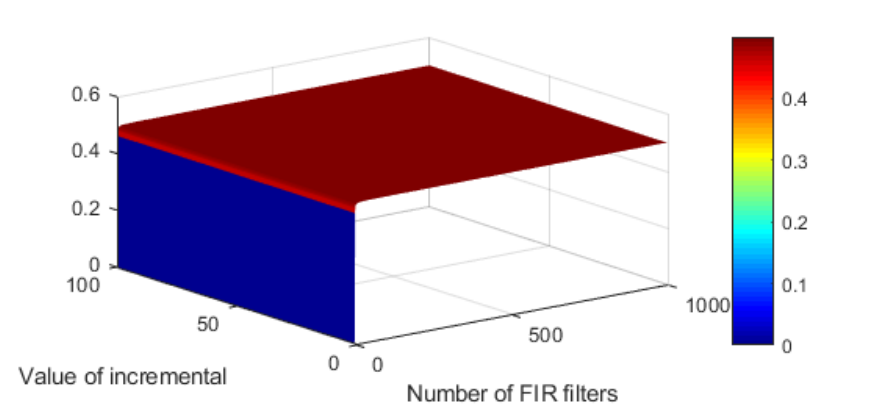}\includegraphics[viewport=0bp 10bp 420bp 210bp,clip,scale=0.6]{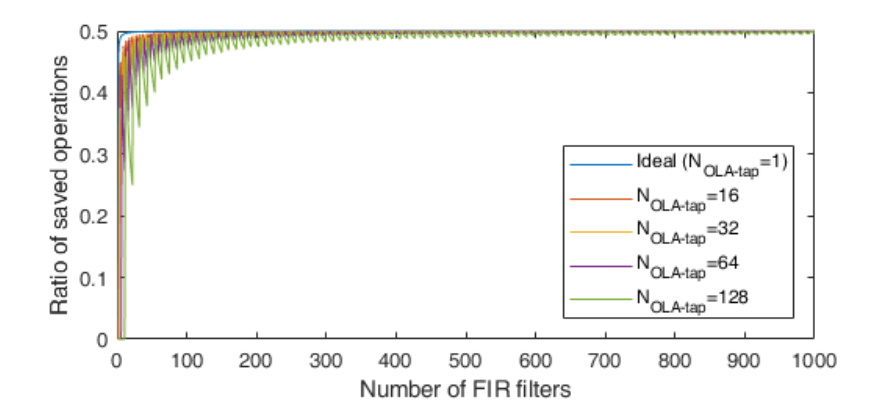}
\par\end{centering}
\begin{centering}
\par\end{centering}
\caption{\label{fig:ola-td-analysis}Reduced taps plotted over tap incremental
and number of filters applying Naive TD (left) and differently sized
OLA algorithms for$N_{inc}=10$, (right)}
\end{figure*}

While the total number of taps that can be saved is a function of
$N_{inc}$ and $N_{filter}$, it can be observed in Figure~\ref{fig:ola-td-analysis}~(left)
that the \emph{ratio} of saved taps to total taps is not affected
by $N_{inc}$. For $N_{inc}=10$, the relationship between the ratio
of saved taps (operations) and the number of FIR filters is presented
in Figure~\ref{fig:ola-td-analysis}~(right). As the number of FIR
filters increases, the ratio of saved taps approaches 0.5, which means
up to 50\% of operations can be saved.

As mentioned before, the OLA algorithm is employed when the number
of taps of a filter is large and not all can be processed at the same
time. Each filter is split into a set of small filters of length $N_{OLA-tap},$see
Figure~\ref{fig:ola-ols}. By employing the OLA algorithm, the total
number of taps is reduced from $Tap_{N_{filter}}N_{filter}$ to 
\[
\sum^{N_{filter}}_{i=1}\left\lceil \frac{N_{tap_{i}}}{N_{OLA-tap}}\right\rceil N_{OLA-tap}.
\]
 This accounts for the fact that $N_{tap_{i}}$ is in general not
perfectly divisible by $N_{OLA-tap}$ and the last small filter needs
to be padded with zeros. In Figure~\ref{fig:ola-td-analysis}~(right)
the smaller the $N_{OLA-tap}$, the larger the ratio of saved operations
and naturally maximised when $N_{OLA-tap}$ is 1. 

\subsubsection{Fourier-Domain~(FD) Filtering}

In a straight-forward FDFIR, the filter is padded to the same length
as the input, whose length in this work is several orders of magnitudes
larger than the filter length. In such an approach, the original filter
length does not influence the execution time for a single filter.
However, the resulting very long Fourier transform makes this approach
inefficient for large input sizes~\cite{wang2019fpga}.

Using instead the OLS algorithm, the input array is split into a set
of chunks, and the length for each chunk is $N_{OLS-FT}$, which is
the Fourier transform length of the OLS algorithm as well. Each chunk
needs to be overlapped with its neighbourhood chunks and the length
of the overlap is the same as the filter length, see Figure~\ref{fig:ola-ols}. 

For an FIR filter $i$, the required number of chunks $N_{OLS-chunk_{N_{i}}}$
is $\left\lceil \frac{N_{input}}{N_{OLS-FT}-Tap_{i}}\right\rceil $.
In generic matched filter group implementation, however, the number
of chunks for every FIR filter is always the same, namely $N_{OLS-chunk_{N_{filter}}}=\left\lceil \frac{N_{input}}{N_{OLS-FT}-Tap_{N_{filter}}}\right\rceil .$
The total number of wasted chunks in a generic implementation is therefore

\[
N_{filter}N_{OLS-chunk_{N_{filter}}}-\sum^{N_{filter}}_{i=1}\left\lceil \frac{N_{input}}{N_{OLS-FT}-Tap_{i}}\right\rceil ,
\]
where $Tap_{N_{filter}}$ has to be smaller than $N_{OLS-FT}$ for
all $N_{filter}$ filters.

For a specific $MF$, the larger the $N_{OLS-FT}$ over $Tap_{N_{filter}}$,
the less unnecessary operations. In a practical implementation, however,
a long FFT engine often performs significantly worse than a short
FFT engine. To search for the most suitable $N_{OLS-FT}$ for a $MF$,
the analysis has to be based on the performance of the specific device,
which will be discussed in Section~\ref{sec:Evaluation}.

\section{\label{sec:Optimistion}Optimisation and Implementation}

\begin{figure*}
\begin{centering}
\includegraphics[viewport=10bp 10bp 550bp 290bp,clip,scale=0.65]{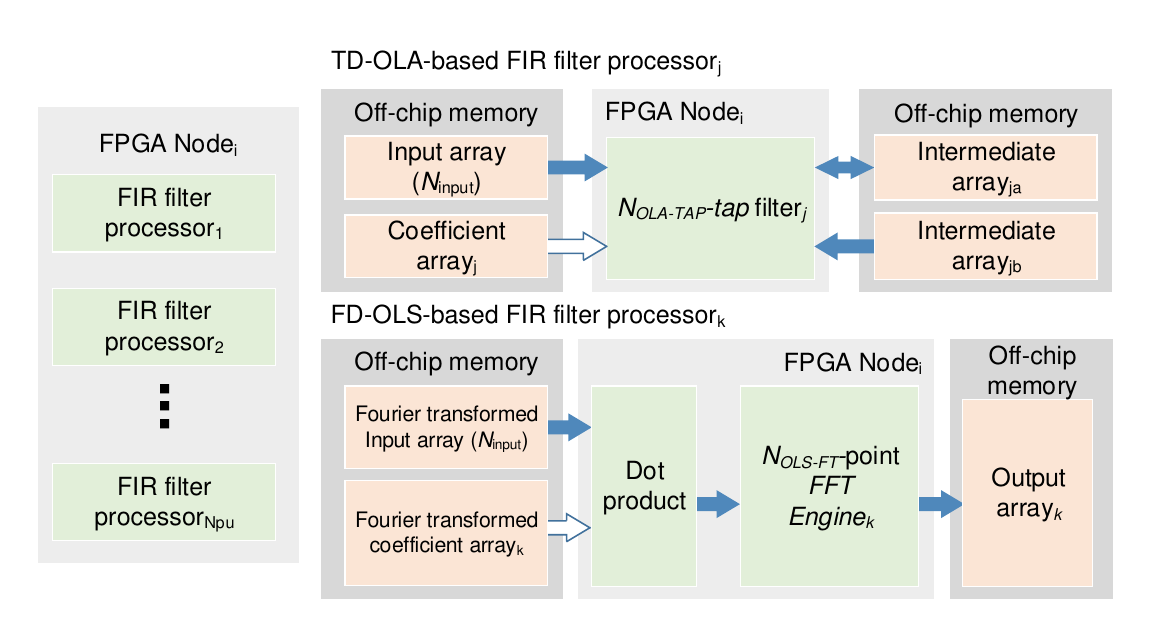}
\par\end{centering}
\caption{\label{fig:Structure-of-a}Architecture of multiple OLA-TD and multiple
OLS-FD implementations on one FPGA node}
\end{figure*}

In this research, it is assumed that the size of input signals is
larger than the device on-chip memory size and the input signals have
to be stored in off-chip memory during processing. The coefficient
array of the filters can be stored in both off-chip memory and on-chip
memory based on its size. The structure of the proposed architectures
is depicted in Figure~\ref{fig:Structure-of-a}, where $N_{pu}$
is the number of processing units. It equals $N_{pu-OLA}$ when employing
TD-OLA and equals $N_{pu-OLS}$ when employing FD-OLS.

\subsection{TD-OLA}

According to Section~\ref{sec:Matched_Filtering}, in TD-OLA based
implementation, FIR filters are assigned to processing units, and
each filter consists of a set of sub-filters. All sub-filters have
the same size, but the number of sub-filters depends on the filter.
Each processing unit works on its sub-filters and sharing the streamed
through input. This repeats unit all filters and their sub-filters
have been processed. This naturally leads to a load balancing problem,
as all processing units should be employed for the entire time. Hence
filters, with their different number of sub-filters, need to be allocated
to the PUs such that each processing unit gets (more or less) an equal
amount of total sub-filters.

\subsubsection{Load Balancing}

The problem of assigning filters to processing units corresponds to
a classical scheduling problem of scheduling independent tasks of
different sizes onto homogeneous processors. This is an NP-hard optimisation
problem, hence no algorithm with polynomial time complexity is know
for the general case. A simple heuristic is to sort the tasks (here
filters) by their size in non-increasing order and to allocate each
task (filter) on the least loaded processor. This heuristic is called
the longest processing time rule (LPT)~\cite{graham1979optimization,graham1969bounds},
which is a 4/3-approximation algorithm, which means that the total
execution time of a solution of this algorithm is not longer than
4/3 times the optimal execution time. 

As per the definition of the filters in Section~\ref{sec:Matched_Filtering},
for any filter $i$ in the matched filter group, where $i\neq N_{filter}$,
$N_{tap_{i}}<N_{tap_{i+1}}$, so the inverse index order is an LPT
order. 

For OLA-TD, the execution time of the $i$th filter contains $\left\lceil \frac{N_{tap_{i}}}{N_{OLA-tap}}\right\rceil $
sub-filters, and the execution latency of each segment is the same. 

\begin{figure*}
\begin{centering}
\includegraphics[viewport=0bp 10bp 943bp 260bp,clip,scale=0.5]{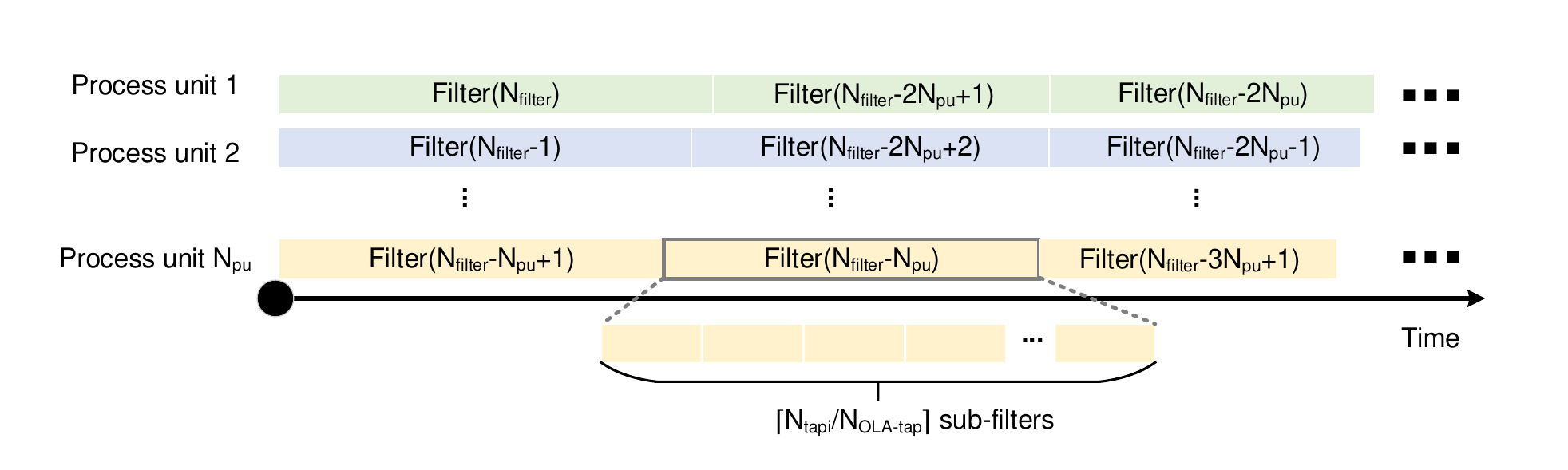}
\par\end{centering}
\caption{\label{fig:lpt}Processing order of TD-OLA-based matched filter group
by employing the LPT rule}
\end{figure*}

\subsubsection{Implementation}

For the TD-OLA algorithm, $N_{pu-OLA}$ filters, whose lengths are
all $N_{OLA-tap}$, are implemented in one FPGA image. The details
of one FIR filter processor using TD-OLA algorithm is depicted in
Figure~\ref{fig:Structure-of-a}. Each processor needs two buffers
in off-chip memory to store the intermediate arrays, whose length
is $2N_{input}$. To avoid simultaneous reading and writing to the
same buffer, which leads to large initiation interval (II), two buffers
are employed.Based on the analysis in Section~\ref{subsec:Theoretical-Analysis},
the smaller the $N_{OLA-tap}$, the less the number of invalid operations
and required logic resources. However, the decrease of $N_{OLA-tap}$
leads to the increase of $N_{pu-OLA}$ in order to perform the same
number of operations simultaneously, because the total number of simultaneous
operations is proportional to $N_{pu-OLA}\times N_{OLA-tap}$. An
increase of $N_{pu-OLA}$ will make the off-chip memory bandwidth
become the main factor that limits the performance as each processing
unit creates a separate output. The number of DSP blocks on an FPGA
and off-chip memory bandwidth are two main factors that affect the
best $N_{pu-OLA}$, especially, when the data types of input signals
and coefficient arrays are floating-point.

For a given data type, the bandwidth needed for each process unit
is fixed. The maximum number of process units that are supported by
a specific device $N_{OLA-omb}$ can be calculated based on the device
off-chip memory bandwidth and the needed bandwidth for each process
unit. The process flow of searching for the optimal $N_{OLA-tap}$
is illustrated in Algorithm~\ref{alg:ola_launch_times}, where $N_{FPGA-tap}$
is the maximum number of taps that can be processed in parallel by
a specific FPGA. 

\begin{algorithm*}
\caption{\label{alg:ola_launch_times}Search for the best tap size $N_{OLA-tap-opt}$
for the TD-OLA based implementation on a specific FPGA}

\textbf{}

\textbf{Require:} $MF-(N_{filter},\,N_{inc},\,Tap_{1})$, $N_{FPGA-tap}$,
$N_{OLA-omb}$.

\textbf{Ensure:} $N_{OLA-omb}\geqslant1$.

1 :~\textbf{~}$Array_{OLA-tap}\leftarrow zero(1,\,N_{FPGA-tap})$
~\{array to store maximum launch times for all $N_{OLA-tap}$ sizes\}

2 :~~\textbf{for} $N_{OLA-tap}=1$ to $N_{FPGA-tap}$\textbf{ do}~\{iterate
over all possible $N_{OLA-tap}$ sizes\}

3 :~~~~~~$N_{pu-OLA}\leftarrow\left\lfloor N_{FPGA-tap}/N_{OLA-tap}\right\rfloor $
~\{number of processing units\}

4 :~\textbf{~}~~~~$Array_{launch}\leftarrow zero(1,\,N_{pu-OLA})$
~\{array to store allocated launch times for this $N_{OLA-tap}$\}

5 :~~~~~~\textbf{if} $N_{pu-OLA}\leqslant N_{OLA-omb}$ \textbf{do}

6 :\textbf{~~}~~~~~~~~\textbf{for $j=N_{filter}$} to $1$
\textbf{do ~}\{use LPT heuristic to allocate filters to processing
units\}

7 :~~~~~~~~~~\textbf{~~}~~$N_{launch-temp}\leftarrow\left\lceil Tap_{j}/N_{OLA-tap}\right\rceil $
~\{launch times for $j$th FIR filter\}

8 :~~~~~~~~~~\textbf{~~~~}$index\leftarrow$get index
of min$(Array_{launch})$ ~\{index of $pu$ with least launch times\}

9 :~~~~~~~~~~~~\textbf{~~}$Array_{launch}(index)\leftarrow Array_{launch}(index)+N_{launch-temp}$
~\{allocate filter to that $pu$\}

10:~~~~~~~~~~\textbf{end for}

11:\textbf{~~}~~~~\textbf{end if} 

12:\textbf{~~}~~~~\textbf{$Array_{OLA-tap}(N_{OLA-tap})\leftarrow$}
max($Array_{launch}$) 

13:~~\textbf{end for}

14:~~\textbf{return }index of $\min(Array_{OLA-tap})>0$ ~\{return
best $N_{OLA-tap-best}$\}
\end{algorithm*}

\begin{figure*}
\begin{centering}
\includegraphics[viewport=0bp 293bp 410bp 553bp,clip,scale=0.6]{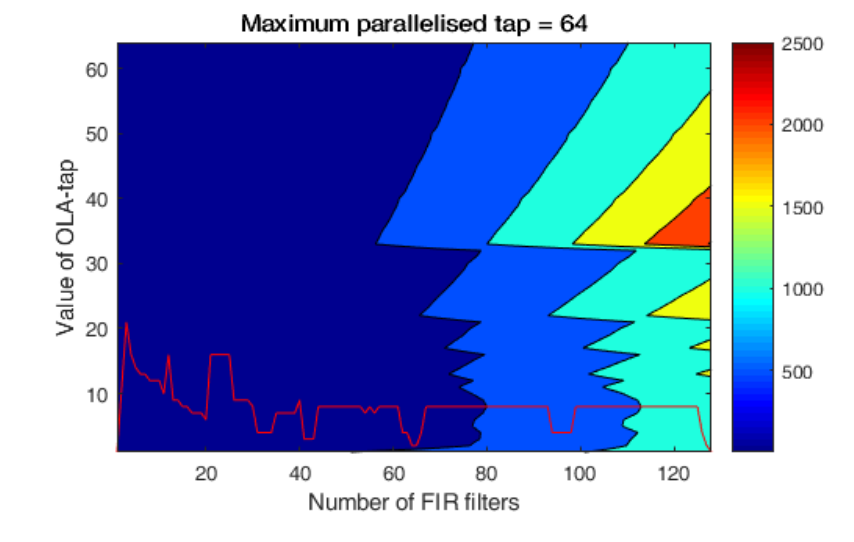}\includegraphics[viewport=0bp 20bp 410bp 280bp,clip,scale=0.6]{ola-launch-times}
\par\end{centering}
\caption{\label{fig:Launch-times-for}Launch times for FPGAs that can parallelise
64 taps (left) and 256 taps (right)}
\end{figure*}

To illustrate the interplay of the number of processing units and
the size of $N_{OLA-tap}$, Figure~\ref{fig:Launch-times-for} depicts
the total number of launch times for an FPGA that can parallelise
64 FIR taps (left, $N_{FPGA-tap}=64$) and 256 taps (right, $N_{FPGA-tap}=256$).
Here, the off-chip memory bandwidth was considered to be not limiting.
The best $N_{OLA-tap}$ for each number of filters $N_{filter}$ in
the group is connected using the red line, which is the index of the
smallest value in each column. For the $N_{FPGA-tap}=64$ FPGA, when
$N_{filter}$ is larger than 14, the $N_{OLA-tap-best}$ is smaller
than 8. For the $N_{FPGA-tap}=256$ FPGA, when $N_{filter}$ is larger
than 30, the $N_{OLA-tap-opt}$ is smaller than 16.

\subsection{FD-OLS}

For FD-OLS based implementation, $N_{pu-OLS}$ FIR filters are processed
in parallel and they all share the same padded input array. The optimisation
problem is to balance the overall size of the padded input arrays
and the total number of processing chunks for a single process unit,
which is to balance the memory cost and processing time as well.

\subsubsection{Filter Size Optimisation}

In order to optimised the FD-OLS design for a matched filter group,
two main factors need to be considered: 1) off-chip memory size and
2) control of loading padded input signals for multiple filters in
parallel\@.

\paragraph{Required off-chip memory}

For the generic FD-OLS implementation (only considering one filter
size, namely $Tap_{N_{filter}}$), the input chunk size for all filters
is the same and the resulting total input size is 
\[
\left\lceil \frac{N_{input}}{N_{OLS-FT}-Tap_{N_{filter}}}\right\rceil N_{OLS-FT}.
\]
 In contrast to the TD-OLA method, where all filters are of different
sizes but the input is the same, here each filter needs a dedicated
padded input chunk and the total input size is increased to
\[
\sum^{N_{filter}}_{i=1}\left\lceil \frac{N_{input}}{N_{OLS-FT}-Tap_{i}}\right\rceil N_{OLS-FT},
\]
 which can even be larger than the outputs of all filters combined,
i.e. the output plane. If $N_{filter}$ is too large, the off-chip
memory size will limit the implementation as not all inputs can fit
there. Hence it will be practical to find a trade-off between different
input sizes to waste less operations and the needed off-chip memory.

\paragraph{Loading control}

Regarding the control of loading, when there are millions of input
signals, thousands of segments have to be processed with each filter
and each length is $N_{OLS-FT}$. Even though the execution latency
of each segment is the same, thousands of segments have to be finished
in one launch instead of processing one or several segments per launch.
The main reason is that the launch overhead is over 10x times slower
than that of Fourier transform $N_{OLS-FT}$ points. Although the
LPT rule could be employed for multiple FD-OLS, the FPGA would have
to load padded input chunks for different filters in parallel and
the complexity of the ensuing loading control system is a problem
that would affect the efficiency of the design.

\begin{figure}
\begin{centering}
\includegraphics[viewport=0bp 400bp 496bp 660bp,clip,scale=0.5]{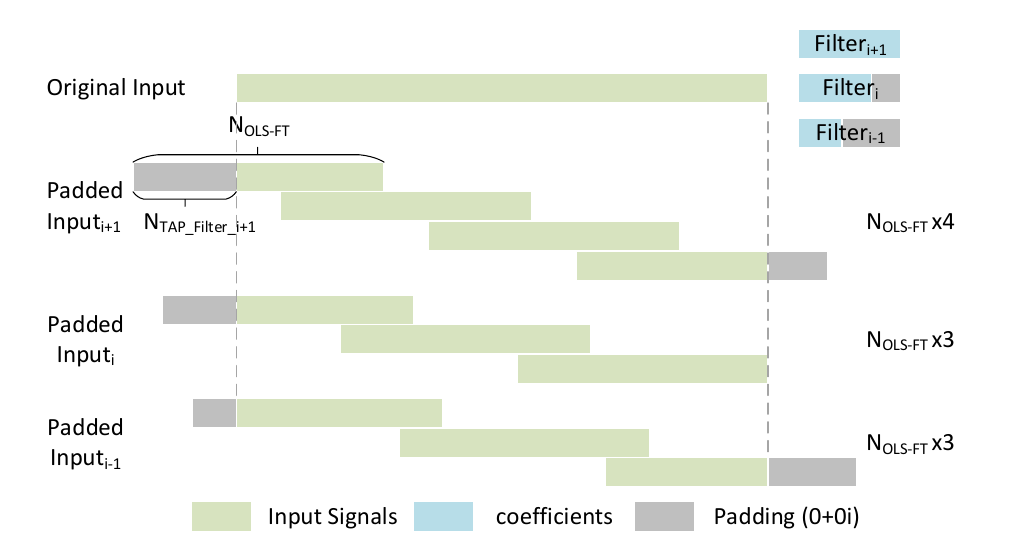} 
\par\end{centering}
\caption{\label{fig:Example-of-padded}Example of padded input group for FD-OLS
method}
\end{figure}

For illustration, look at an example with three filters (filter$_{i-1}$,
filter$_{i}$, and filter$_{i+1}$) in Figure~\ref{fig:Example-of-padded}.
The padded input$_{i+1}$ (following the OLS method, see Figure~\ref{fig:ola-ols})
is compatible with filters whose length is shorter than $N_{tap_{i+1}}$,
such as filter$_{i}$ and filter$_{i-1}$, but the padded input$_{i-1}$
only fits for filter$_{i-1}$. One disadvantage regarding the number
of chunks in Figure~\ref{fig:Example-of-padded} is that the needed
space for filter$_{i}$ and filter$_{i-1}$ are both $3N_{OLS-FT}$,
but it is $4N_{OLS-FT}$ for filter$_{i+1}$. More chunks means consuming
more off-chip memory space and longer execution time. However, a dedicated
padded input for each filter costs more off-chip memory. 

As a trade-off, in this research, we propose the method that $N_{share}N_{pu-OLS}$
filters share the same padded input chunks, where $N_{pu-OLS}$ is
the number of processing units and $N_{share}$ is an integer constant
and $N_{share}\geqslant1$. For each $N_{share}N_{pu-OLS}$ filters,
i.e. a sub-group, the lengths of all filters are padded to the same
as the longest filter in these $N_{share}N_{pu-OLS}$ filters. In
this case, the number of chunks for each process unit is the same
and it is unnecessary to load different padded input chunks for each
process unit, one chunk is shared by all processing units.

\subsubsection{Implementation}

The architecture of the FPGA-based implementation is given in Figure~\ref{fig:Structure-of-a}.
For a single processor, it consists of two parts: dot product and
$N_{OLS-FT}$-point FFT engine. It has two significant differences
to the TD-OLA implementation: 1)~the input signals and coefficient
arrays have to be Fourier transformed before processing, and 2)~the
output from the FFT engine can be sent directly to off-chip memory
without loading and adding the output from the previous block (as
necessary with OLS).

In this section, we investigate the relationship between the size
of the padded input chunks and the execution time of the FD-OLS based
implementation. The number of processing units $N_{pu-OLS}$ on a
specific FPGA is restricted by the number of available logic resources
and it will be evaluated in Section~\ref{subsec:Resource-Usage}.
When $N_{share}N_{pu-OLS}$ filters, which is a sub-group, share the
same padded input chunks, there will be $\left\lceil N_{filter}/(N_{share}N_{pu-OLS})\right\rceil $
different padded input sets in total. The size of each padded input
set is different and each of them is larger than that of original
input data.

There are two main factors that influence the execution times: 1)~the
length of the overlapped part ($Tap_{i}$) and 2)~the number of chunks
for each sub-group 
\[
N_{OLS-chunk_{i}}=\left\lceil \frac{N_{input}}{N_{OLS-FT}-Tap_{i}}\right\rceil 
\]
, where FIR filter $i$ is the largest filter in the sub-group of
filters.. These two factors for all $N_{inputgroup}$ filter sub-groups,
where $N_{inoutgroup}=N_{filter}/N_{share}N_{pu-OLS}$, are not the
same and these changes can be implemented in the host programs, which
is processed by CPU. The overlap can be simply created using memory
copying, e.g. using the \texttt{memcpy} function in C/C++. Regarding
the second factor, this just changes the iteration bounds of the loop.

The only parameter that can be adjusted for a given $MF$ on a specific
FPGA is $N_{share}$. Based on requirements such as execution time
or memory space, the best $N_{share}$ is different. Two algorithms
are proposed, a time-efficient one, minimizing the execution time
and an area-efficient one, minimising the required memory space. The
details of time-efficient $N_{share}$ are in Algorithm~\ref{alg:time-efficient Nshare}
and area-efficient $N_{share}$ are in Algorithm~\ref{alg:area-efficient Nshare}.

\begin{algorithm*}
\caption{\label{alg:time-efficient Nshare}Searching for the best time-efficient
$N_{share}$}

\textbf{Require:} $MF-(N_{filter},\,N_{inc},\,Tap_{1})$, $N_{input}$,
$N_{pu-OLS}$, $N_{OLS-FT}$.

\textbf{Ensure:} $N_{OLS-FT}\geqslant Tap_{N_{filter}}$, $N_{share}N_{pu-OLS}\leqslant N_{filter}$.

1 :~~$N_{share-max}\leftarrow\left\lceil N_{filter}/N_{pu-OLS}\right\rceil $~\{maximum
$N_{share}$\}

2 :~~$Array_{totalchunk}\leftarrow zero(N_{share-max})$ ~\{array
of number of chunks for each $N_{share}$\}

3 :~~\textbf{for $N_{share}=1$} to $N_{share-max}$ \textbf{do}

4 :~~~~~~$N_{inputgroup}\leftarrow\left\lceil N_{filter}/(N_{share}N_{pu-OLS})\right\rceil $
~\{number of filter sub-groups\}

5 :~~\textbf{~~~~for $j=1$} to $N_{inputgroup}$ \textbf{do}

6 :~~\textbf{~~~~}~~~~$N_{chunk-temp}\leftarrow N_{share}\left\lceil N_{input}/(N_{OLS-FT}-Tap_{N_{filter}-(j-1)N_{share}N_{pu-OLS}})\right\rceil $ 

~~\textbf{~~~~}~~~~~~\textbf{~}\{number of input chunks
of $jth$ filter sub-group\}

7 :~~\textbf{~~~~}~~~~$Array_{totalchunk}(N_{share})\leftarrow Array_{totalchunk}(N_{share})+N_{chunk-temp}$
~\{total number of chunks for current $N_{share}$\}

8 :\textbf{~~~~~~end for} 

9 :~~\textbf{end for}

10:~~\textbf{return }index of min($Array_{totalchunk}$) ~\{best
time-efficient $N_{share}$\} 
\end{algorithm*}

\begin{algorithm*}
\caption{\label{alg:area-efficient Nshare}Searching for the best area-efficient
$N_{share}$}

\textbf{Require:} $MF-(N_{filter},\,N_{inc},\,Tap_{1})$, $N_{input}$,
$N_{pu-OLS}$, $N_{OLS-FT}$.

\textbf{Ensure:} $N_{OLS-FT}\geqslant Tap_{N_{filter}}$, $N_{share}N_{pu-OLS}\leqslant N_{filter}$.

1 :~~$N_{share-max}\leftarrow\left\lceil N_{filter}/N_{pu-OLS}\right\rceil $~\{maximum
$N_{share}$\}

2 :~~$Array_{totalsize}\leftarrow zero(N_{share-max})$ ~\{array
of number of chunks for each $N_{share}$\}

3 :~~\textbf{for $N_{share}=1$} to $N_{share-max}$ \textbf{do}

4 :~~~~~~$N_{inputgroup}\leftarrow\left\lceil N_{filter}/(N_{share}N_{pu-OLS})\right\rceil $
~\{number of filter sub-groups\}

5 :~~~~~~$N_{size-temp}\leftarrow0$

6 :~~\textbf{~~~~for $j=1$} to $(N_{inputgroup}-1)$ \textbf{do}

7 :~~\textbf{~~~~}~~~~$N_{size-temp}\leftarrow N_{size-temp}+(1+N_{share}N_{pu-OLS})\left\lceil N_{input}/(N_{OLS-FT}-Tap_{N_{filter}-(j-1)N_{share}N_{pu-OLS}})\right\rceil $ 

~~\textbf{~~~~}~~~~~~\textbf{~}\{number of input and
output chunks of $jth$ filter sub-group\}

8 :\textbf{~~~~~~end for} 

9 :~~~~~~$N_{size-temp}\leftarrow N_{size-temp}+(N_{share}N_{inputgroup}N_{pu-OLS}-N_{filter}+1)\times$

~~~~~~~~~$\left\lceil N_{input}/(N_{OLS-FT}-Tap_{N_{filter}-(N_{inputgroup}-1)N_{share}N_{pu-OLS}})\right\rceil $ 

10:~~~~~~$Array_{totalsize}(N_{share})\leftarrow Array_{totalsize}(N_{share})+N_{size-temp}$

11:~~\textbf{end for}

12:~~\textbf{return }index of min($Array_{totalsize}$) ~\{return
best area-efficient $N_{share}$\} 
\end{algorithm*}

\begin{figure}
\begin{centering}
\includegraphics[viewport=0bp 0bp 280bp 220bp,clip,scale=0.8]{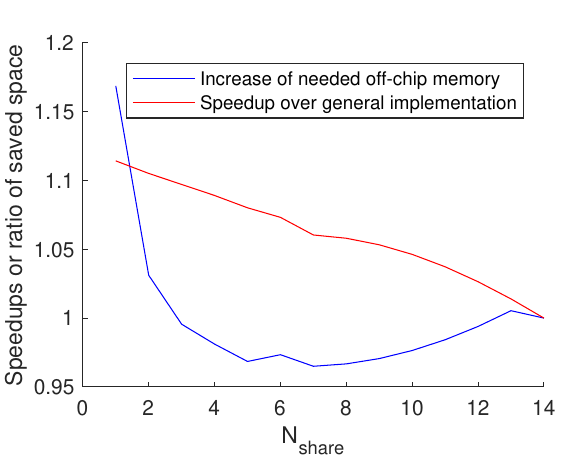}
\par\end{centering}
\caption{\label{fig:Speedup-and-saved}Speedup and saved memory space by changing
the number of shared padded inputs through $N_{share}$ }
\end{figure}

Taking the $MF$ in~\cite{wang2019fpga} as an example, which is
$MF-(42,\,10,\,1)$, and the input signal is $2^{21}$ values long,
the number of process units is $N_{pu-OLS}=3$, and the FFT length
with the best performance is $N_{OLS-FT}=2,048$. The proposed method
is compared with the generic implementation in~\cite{wang2019fpga}.
The speedup in terms of execution latency and the increase of needed
off-chip memory size are given in Figure~\ref{fig:Speedup-and-saved}.
As can be seen, the speedup is the highest when every three filters,
i.e. $N_{share}=1$, share one padded input, which is 1.12x faster.
However, it costs about 15\% more memory space. When 21 filters share
one padded input, i.e. $N_{share}=7$, the speedup is only 1.06x,
but up to 4\% memory space can be saved. In the generic implementation,
all 42 filters share one padded input, which corresponds to $N_{share}=14$
in Figure~\ref{fig:Speedup-and-saved}.

\section{\label{sec:Evaluation}Experiments and Results}

In this section we perform various experimental evaluations of the
proposed approaches. Our objective is i) to evaluate the applicability
of the approaches on different FPGA platforms, ii) to determine the
best performing parameters for the different approaches, iii) to demonstrate
the benefit of the optimised approaches over the generic implementation
iv) to compare the performance with GPU-optimised implementations. 

\subsection{Experimental Setup}

Four types of FPGA devices are employed in the experimental evaluation
, which are Terasic DE5 board with Intel Stratix V, referred to as
$\mathbf{S5}$, Nallatech 385A card with Intel Arria 10, referred
to as $\mathbf{A10}$, Intel Xeon Scalable processor with an in-package
Arria 10, referred to as $\mathbf{X+A10}$, and Xilinx VCU1525 with
Virtex UltraScale+, referred to as $\mathbf{U9}$. Their technical
specifications are given in Table~\ref{tab:Details-of-FPGA}. X+A10
is from the Intel Hardware Accelerator Research Program (HARP), and
the FPGA is connected with the CPU through both the PCI express~(PCIe)
bus and the QPI bus. The S5, A10, and U9 cards are connected to the
host processor through the PCIe bus.

\begin{table*}
\caption{\label{tab:Details-of-FPGA}Specifications of the employed FPGA platforms}

\begin{centering}
{\small}{\small\par}
\par\end{centering}
\centering{}%
\begin{tabular}{|c|c|c|c|c|}
\hline 
\multirow{2}{*}{Device} & Terasic DE5-Net  & Nallatech 385A  & Xeon+FPGA & Xilinx VCU1525\tabularnewline
 & (\textbf{S5}) & (\textbf{A10}) & (\textbf{X+A10}) & (\textbf{U9})\tabularnewline
\hline 
\hline 
\multirow{2}{*}{Hardware} & Intel Stratix V  & Intel Arria 10  & Intel Arria 10  & Xilinx Virtex UltraScale+\tabularnewline
 & 5SGXA7 & GX1150 & GX1150 & XCVU9P\tabularnewline
\hline 
Technology & $28nm$ & $20nm$ & $20nm$ & $16nm$\tabularnewline
\hline 
\multirow{2}{*}{Compute resource} & 622,000 LEs & 1,506,000 LEs & 1,506,000 LEs & 2,586,000 LEs\tabularnewline
\cline{2-5}
 & 256 DSP blocks & 1,518 DSP blocks & 1,518 DSP blocks & 6,840 DSP blocks\tabularnewline
\hline 
On-chip memory size & 50$Mb$ & 53$Mb$ & 53$Mb$ & 345.9 $Mb$\tabularnewline
\hline 
Off-chip memory size & 2 x $2GB$ DDR3 & 2 x $4GB$ DDR3 & \textendash{} & 4 x $16GB$ DDR4\tabularnewline
\hline 
OpenCL global & \multirow{2}{*}{$25,600MB/s$} & \multirow{2}{*}{$35,128MB/s$} & \multirow{2}{*}{$25,000MB/s$} & \multirow{2}{*}{}\tabularnewline
memory bandwidth &  &  &  & \tabularnewline
\hline 
Max clock frequency & 600$MHz$ & 1.5$GHz$ & 1.5$GHz$ & \tabularnewline
\hline 
Max power consumption & \textemdash{} & 75W & \textendash{} & 225W\tabularnewline
\hline 
\end{tabular}
\end{table*}

For different FPGA series and devices, the high-level synthesis approach
OpenCL~\cite{munshi2009opencl} is employed to enable the fast coverage
of different design approaches and design space parameters. All Intel
FPGA-based OpenCL kernels are compiled using the Intel OpenCL offline
compiler~(AOC) version 16.0.0.211. For Xilinx UltraScale+ FPGA based
OpenCL kernels, they are compiled with Xilinx SDx for OpenCL (XOCC)
version 2018.2. 

All evaluated kernels on Intel FPGAs are the same and they are compiled
with the same commands. Essentially the same kernel codes are evaluated
on Xilinx FPGA, with differences in the compiler directives and optimizations.
The used optimization features for both compiler are given in Table~\ref{tab:Optimization-Features}.
The main difference is that Xilinx does not provide optimization features
for floating-point operations.

\begin{table*}
\caption{\label{tab:Optimization-Features}Optimization features of Intel FPGA-based
and Xilinx-based OpenCL development}

\centering{}%
\begin{tabular}{|c|c|c|}
\hline 
Optimization Features & Intel (AOC) & Xilinx (XOCC)\tabularnewline
\hline 
\hline 
Pipe & Channel and Pipe & Pipe\tabularnewline
\hline 
Loop unroll & \#pragma unroll & \_\_attribute\_\_((opencl\_unroll\_hint))\tabularnewline
\hline 
\multirow{2}{*}{Pipelining} & \multirow{2}{*}{Automatically} & \_\_attribute\_\_((xcl\_pipeline\_loop))\tabularnewline
 &  & \_\_attribute\_\_((xcl\_dataflow))\tabularnewline
\hline 
Kernel Vectorization & \_\_attribute\_\_((num\_simd\_work\_item)) & \_\_attribute\_\_((vec\_type\_hint))\tabularnewline
\hline 
Floating-point  & balanced tree (\textendash fp-relaxed) & \multirow{2}{*}{No related attributes}\tabularnewline
operations & rounding operations (\textendash fpc) & \tabularnewline
\hline 
Multiple memory banks & -no-interleaving=<global\_memory\_type> & -{}-max\_memory\_ports \textendash sp ....\tabularnewline
\hline 
Local memory partition & \_\_attribute\_\_((<memory\_type>, ...)) & \_\_attribute\_\_((xcl\_array\_partition))\tabularnewline
\hline 
\end{tabular}
\end{table*}

\subsection{\label{subsec:Resource-Usage}Resource Usage and Performance}

In this section, we evaluate the resource usage and performance of
the multiple TD-OLA-($N_{OLA-tap}$) structure and the multiple FD-OLS-($N_{OLS-FT}$)
structure on employed FPGAs. The basic single TDFIR is evaluated before
investigating multiple OLA-TD and the basic single FDFIR is evaluated
before investigating multiple OLS-FD. The data types of input signals
and coefficients are both complex single-precision floating point.
The performance in terms of $GFLOPS$ of the FD designs is not compared
with the results of TD designs, because due to the very different
approaches and the required total computations it is not meaningful
to do that.

\subsubsection{TD-OLA \textendash{} Single}

\begin{figure*}
\begin{centering}
\includegraphics[viewport=0bp 10bp 410bp 310bp,clip,scale=0.6]{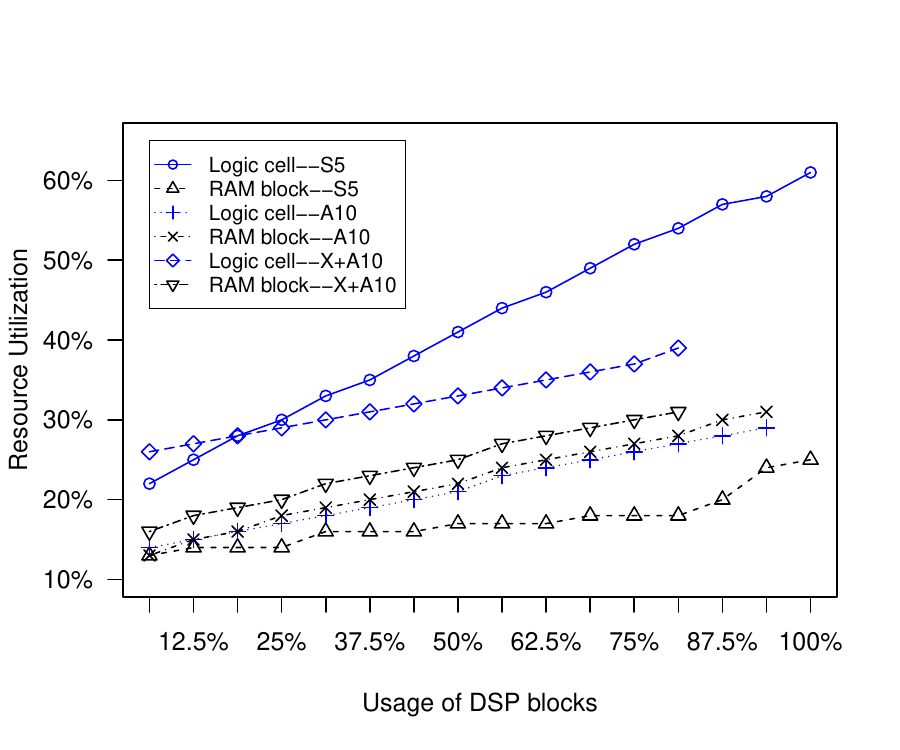}\includegraphics[viewport=0bp 10bp 410bp 310bp,clip,scale=0.6]{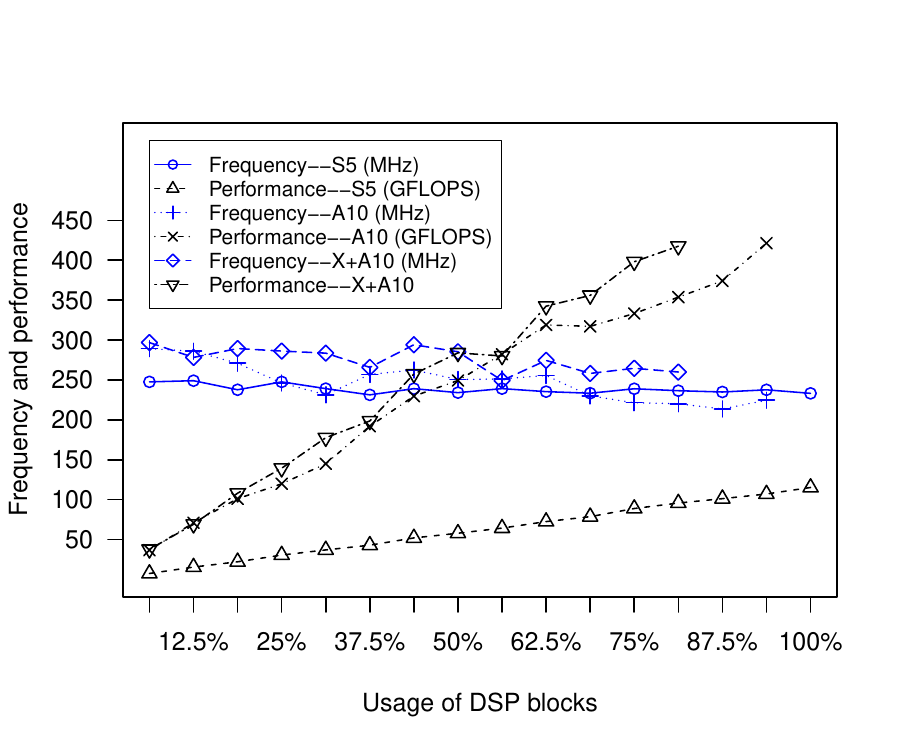}
\par\end{centering}
\caption{\label{fig:tdfir_resource}Resource usage (left) and performance (right)
over DSP block usage}
\end{figure*}

The usage of DSP block plays an important role in implementing TD
filters and it is the first resource type that is used up as the the
number parallelised TD taps is increased. The change of resource usages
in terms of logic cells and RAM blocks is therefore plotted over the
usage of DSP blocks in Figure~\ref{fig:tdfir_resource}(left) and
the kernel frequencies and performance in $GFLOPS$ are given in Figure~\ref{fig:tdfir_resource}
(right).

For A10 and X+A10, the usage of logic cells and RAM blocks are less
than 40\% for an DSP block usage of over 80\%. The kernel frequency
for all implementations ranges from $200MHz$ to $300MHz$ and decreases
as DSP block usage increases. The maximum performance of A10 and X+A10
are around $420GFLOPS$, and $120GFLOPS$ for S5. Although A10 and
X+A10 have enough DSP blocks to parallelise a 256-tap filter, the
AOC compiler cannot generate the bitstream file when the length is
larger than 240 on A10 and 208 on X+A10.

For the Xilinx FPGA, the compilation period for kernels with floating
point operations is a barrier for the efficient employment for TD
filters. Already for an FIR with 128 parallelised -taps, the compilation
time is up to 10 hours. For longer FIR filters such as a 256-tap filter,
the compiler does not finish within 24 hours. A possible explanation
is that Xilinx FPGAs and XOCC do not have dedicated optimisations
for floating point operations and a lack of floating point hardware
blocks. The performance of a 64-tap filter and a 128-tap filter on
U9 are $120GFLOPS$ and $244GFLOPS$, respectively, and the kernel
frequencies are both $233MHz$.

To further explore the behaviour of the U9, we change the data type
to integer (32-bit), for which the bitstream file can be generated
in several hours and the corresponding results are given in Table~\ref{tab:basic_u9}.
For a 256-tap filter, the kernel frequency is only $140MHz$ and the
performance is $252GOPS$. 

\begin{table}
\caption{\label{tab:basic_u9}Details of basic TDFIR using Xilinx U9}

\centering{}%
\begin{tabular}{|c|c|c|c|c|}
\hline 
Filter  & DSP block & Frequency & Performance & II\tabularnewline
taps & \% & ($MHz$) & ($GOPS$, 32-bit) & \tabularnewline
\hline 
\hline 
\multirow{1}{*}{16} & 3.8 & 250 & 14 & 1\tabularnewline
\hline 
\multirow{1}{*}{64} & 15.1 & 275 & 123 & 1\tabularnewline
\hline 
128 & 30.1 & 240 & 207 & 1\tabularnewline
\hline 
256 & 60.2 & 140 & 252 & 1\tabularnewline
\hline 
\end{tabular}
\end{table}

\subsubsection{TD-OLA \textendash{} Multiple}

Now we evaluate multiple TD-OLA filters, informed by the resource
consumption limits measured in the previous experiments.There are
two parameters that can be varied: the number of parallelised taps
per filter, $N_{OLA-tap}$ and the number of parallel processing units
$N_{pu-OLA}$. The aim is to select combinations of these two parameters
which (almost) exhaust the FPGA resources. Results are given in Table~\ref{tab:Process-units-for}.

As recommended by FPGA best practice guides, all considered $N_{OLA-tap}$
values are powers of two~\cite{altera2016openclpra}. For each such
value we determine the maximum number of processing units $N_{pu-OLA}$
that still fits on the FPGA. This implies that compilation was not
successful for larger $N_{pu-OLA}$(e.g. $N_{pu-OLA}=2$ for $N_{OLA-tap}$
did not successfully compile).

For both S5 and A10, the kernel frequency drops as $N_{pu-OLA}$ increases.
As can be seen the usage of DSP block is proportional to the the product
of $N_{pu-OLA}$ and $N_{OLA-tap}$. However, the usage of logic cell
and RAM block increases with $N_{pu-OLA}$.

\begin{table*}
\caption{\label{tab:Process-units-for}Process units for algorithms on a specific
device}

\begin{centering}
{\scriptsize}{\scriptsize\par}
\par\end{centering}
\centering{}%
\begin{tabular}{|c|c|c|c|c|c|c|c|c|}
\hline 
\multicolumn{1}{|c|}{Device} & \multicolumn{4}{c|}{S5} & \multicolumn{4}{c|}{A10}\tabularnewline
\hline 
\hline 
$N_{OLA-tap}$ & 8 & 16 & 32 & 64 & 16 & 32 & 64 & 128\tabularnewline
\hline 
$N_{pu-OLA}$ & 8 & 4  & 2 & 1 & 15 & 7 & 3 & 1 \tabularnewline
\hline 
Logic cells (\%) & 71 & 68 & 64 & 50 & 47 & 33 & 25 & 21\tabularnewline
\hline 
DSP block (\%) & 100 & 100 & 100 & 100 & 94 & 88 & 76 & 50\tabularnewline
\hline 
RAM block (\%) & 38 & 32 & 26 & 20 & 64 & 35 & 23 & 17\tabularnewline
\hline 
Frequency ($MHz$) & 212 & 223 & 226 & 232 & 153 & 191 & 213 & 258\tabularnewline
\hline 
Bits/clock cycle & 576 & 320 & 192 & 128 & 1,024 & 512 & 256 & 128\tabularnewline
\hline 
Bandwidth ($MBytes/s$) & 15,264 & 8,920 & 5,424 & 3,712 & 19,584 & 12,224 & 6,816 & 4,128\tabularnewline
\hline 
\end{tabular}
\end{table*}

\begin{figure*}
\begin{centering}
\includegraphics[viewport=0bp 0bp 280bp 220bp,clip,scale=0.8]{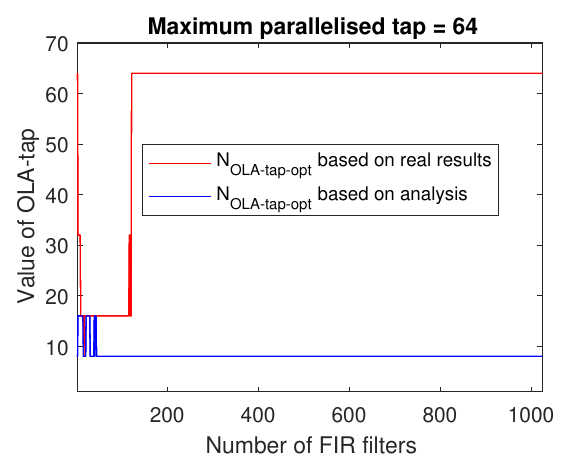}\includegraphics[viewport=0bp 0bp 280bp 220bp,clip,scale=0.8]{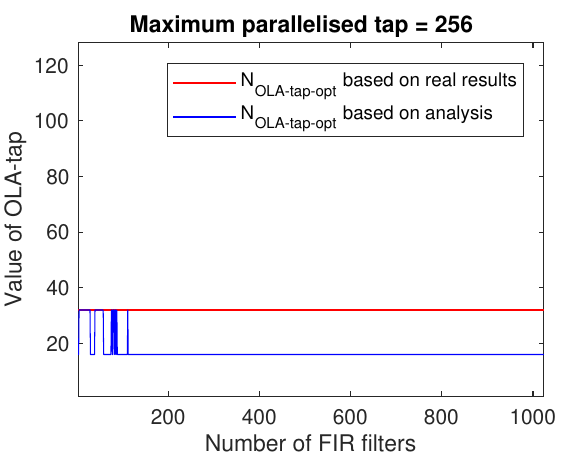}
\par\end{centering}
\caption{\label{fig:-real_Nopt}$N_{OLA-tap-opt}$ based on evaluation of implementations
for S5 (left) and A10 (right) }
\end{figure*}

When considering the required bandwidth (in stream mode) for the different
parameter combinations, we observe for all designs that it is smaller
than the maximum OpenCL global memory bandwidth as listed in Table~\ref{tab:Details-of-FPGA}.
In this case, the device off-chip memory bandwidth is not a performance
barrier for most of $MF$. In Figure~\ref{fig:Launch-times-for}
we studied analytically the value of $N_{OLA-tap}$ over the number
of filters. As the frequency of all designs varies slightly in real
implementations and the $N_{pu-OLA}$ might be different from the
theoretical analysis in Figure~\ref{fig:Launch-times-for} for some
designs due to the fact that some values cannot be compiled/synthesised
(e.g. A10 cannot parallelise 256-taps), the best valueof $N_{OLA-tap}$
for some $MF$s might change.Details are depicted in Figure~\ref{fig:-real_Nopt},
based on results of Table~\ref{tab:Process-units-for}, where the
blue line is showing the values from Figure~\ref{fig:Launch-times-for}.
On S5, $N_{OLA-tap-opt}$ increases from 8 to 16 for $MF$s with less
than 160 filters and 64 for $MF$s with more than 160 filters. In
terms of the $N_{OLA-tap-opt}$ on A10, it is increased from 16 to
32 for most of the $MF$s.

\subsubsection{\label{subsec:FD-OLS-=002013-FFT}FD-OLS \textendash{} FFT}

For the FD-OLS implementation (as can be seen in Figure~\ref{fig:Structure-of-a}),
the central element is the FFT engine, and we evaluate the resource
usage and performance of the FFT engine in this section. Three types
of FFT engines are evaluated, which are 4-point FFT engine, 8-point
FFT engine, and 16-point FFT engines. An x-point FFT engine can process
x points in parallel.

\begin{figure*}
\begin{centering}
\includegraphics[viewport=0bp 10bp 410bp 310bp,clip,scale=0.6]{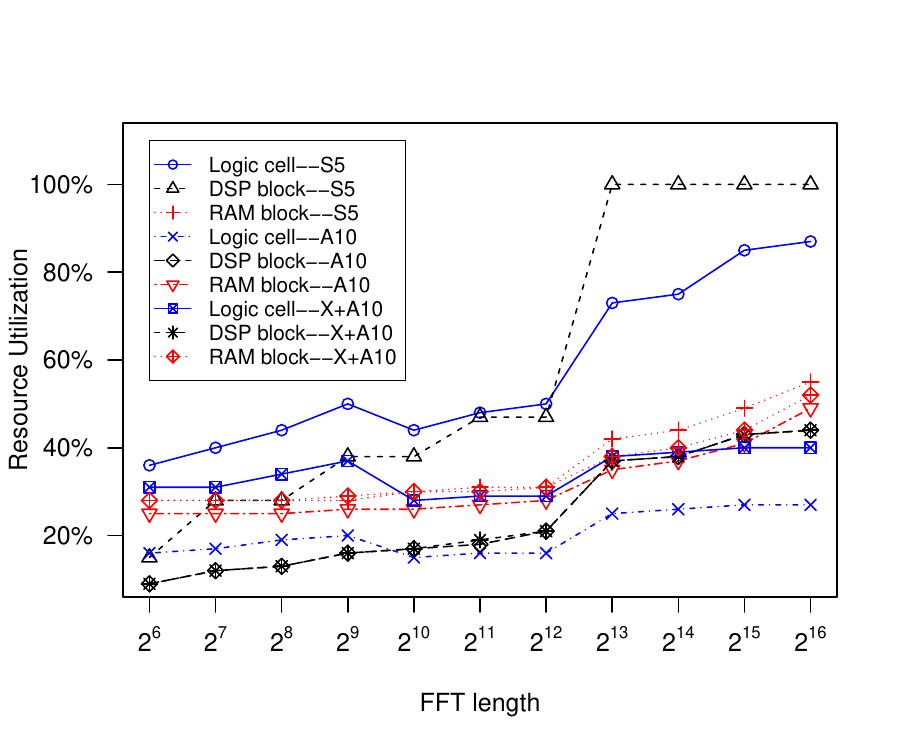}\includegraphics[viewport=0bp 10bp 410bp 310bp,clip,scale=0.6]{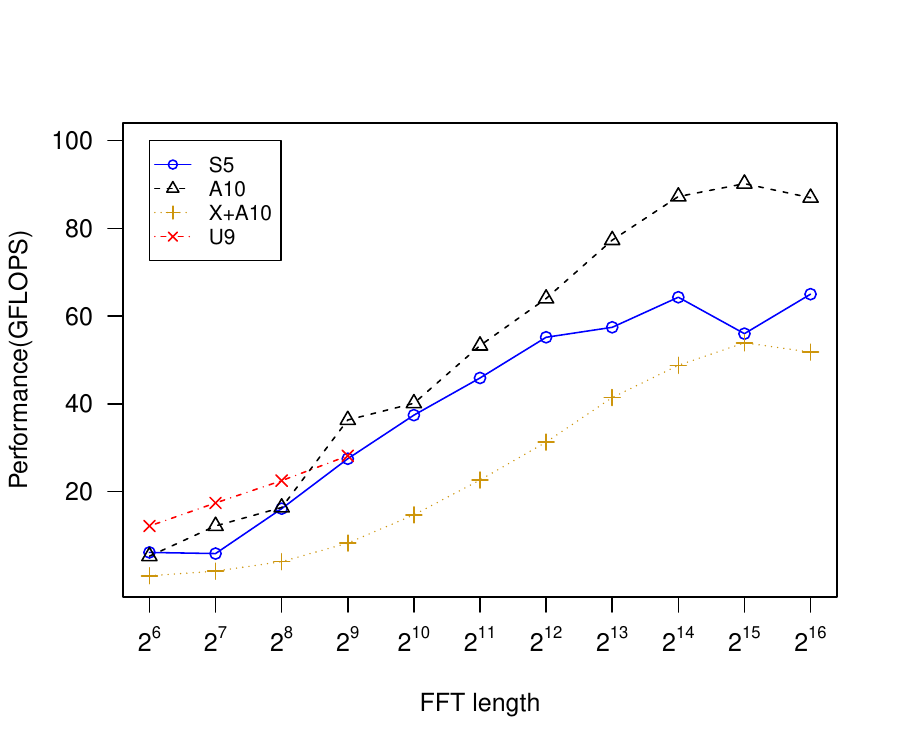}
\par\end{centering}
\caption{\label{fig:Resource_performance}Resource usage (left) and performance
(right) of 8-point FFT engine}
\end{figure*}

The resource usage and performance of 8-point engine are presented
in Figure~\ref{fig:Resource_performance}. The pipelined 8-point
FFT engine is evaluated on all FPGA platforms (Table~\ref{tab:Details-of-FPGA})
including U9. On the U9, the evaluated length ranges from $2^{6}$
to $2^{9}$ only. When the FFT length is larger than $2^{10}$, the
array for processing cannot be instantiated using registers and the
II becomes larger than 1, loop pipelining cannot be achieved. Besides
this issue, the input signals have to be reordered to achieve stream
mode on U9. In general, it can be seen that there is an increase for
all types of resources (Logic cell, DSP block, and RAM block) when
the FFT length grows from $2^{12}$ to $\geq2^{13}$. Regarding the
logic cells, their usage for all devices drops when the FFT length
increases from $2^{9}$ to $2^{10}$.

\begin{figure*}
\begin{centering}
\includegraphics[viewport=0bp 10bp 410bp 310bp,clip,scale=0.6]{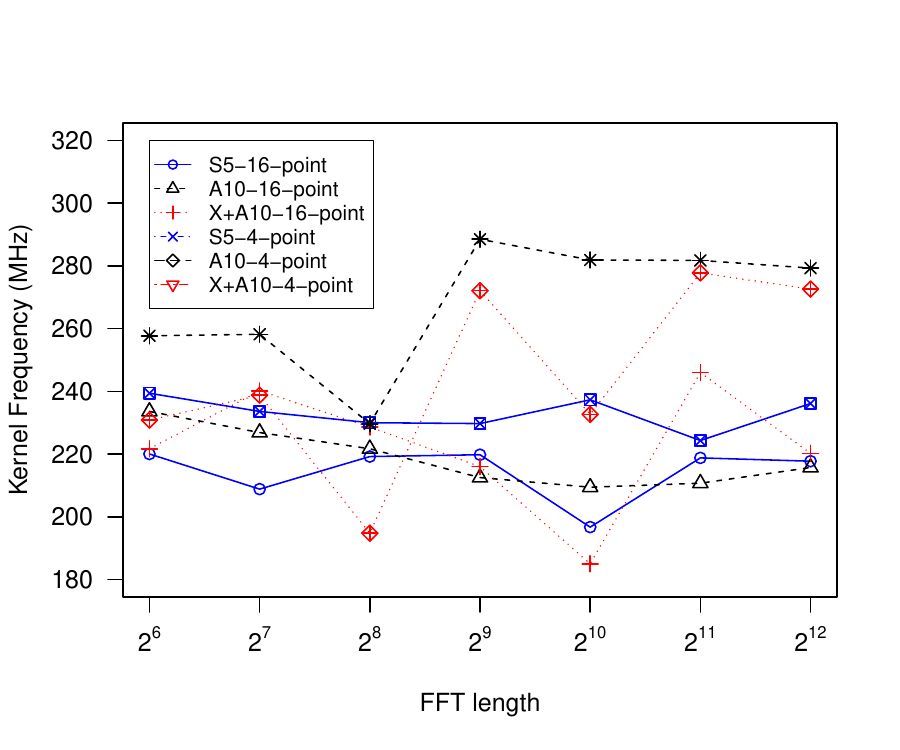}\includegraphics[viewport=0bp 10bp 410bp 310bp,clip,scale=0.6]{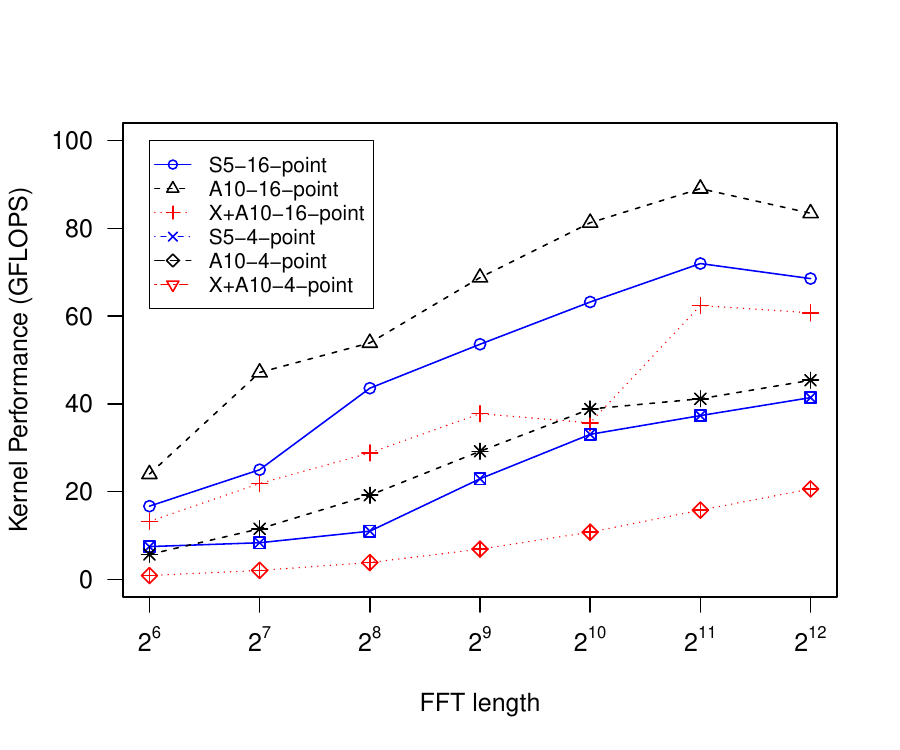}
\par\end{centering}
\caption{\label{fig:Freq_perf_4_16_FFT}Frequency~(left) and performance~(right)
of 4-point and 16-point FFT engines}
\end{figure*}

The frequency and performance of 4-point and 16-point FFT engines
are given in Figure~\ref{fig:Freq_perf_4_16_FFT}. For most of the
kernels, the frequency ranges from $180MHz$ to $300MHz$. Except
for $2^{10}$ 16-point FFT engine on X+A10, the performance of 16-point
FFT engine is higher than that of 4-point FFT engine. The A10 performs
best and among these Intel devices and S5 performs better than X+A10.

\subsubsection{FD-OLS \textendash{} Multiple}

Using the performance and resource usage results of a single FFT engine
in Section~\ref{subsec:FD-OLS-=002013-FFT}, we evaluate now multiple
OLS-FD filters. First, we determine the maximum number of processing
units $N_{pu-OLS}$ that can be instantiated without breaching the
maximum bandwidth limit of the targetted FPGA device. The required
off-chip memory bandwidth for the three types of FFT engines is shown
in Table~\ref{tab:Required-off-chip-memory}, where all FFT engines
share the same padded inputs. The required bandwidth is calculated
on the assumption that the kernel can achieve stream mode and is running
at $200MHz$. If the frequency is higher than $200MHz$, the required
bandwidth increases as well. The maximum $N_{pu-OLS}$ for a specific
device can then be determined from the the device's maximum OpenCL
global memory bandwidth given in Table~\ref{tab:Details-of-FPGA}.

\begin{table*}
\caption{\label{tab:Required-off-chip-memory}Required off-chip memory bandwidth
for an FFT engine}

\centering{}%
\begin{tabular}{|c|c|c|c|c|c|c|c|}
\hline 
\multicolumn{2}{|c|}{FFT engine } & \multicolumn{2}{c|}{4-point} & \multicolumn{2}{c|}{8-point} & \multicolumn{2}{c|}{16-point}\tabularnewline
\hline 
\hline 
\multicolumn{2}{|c|}{} & $1st$ & $+1$ & $1st$ & $+1$ & $1st$ & $+1$\tabularnewline
\hline 
\multicolumn{2}{|c|}{Bits/clock cycle} & 512 & 256 & 1,024 & 512 & 2,048 & 1,024\tabularnewline
\hline 
\multicolumn{2}{|c|}{Required bandwidth } & \multirow{2}{*}{12,800} & \multirow{2}{*}{6,400} & \multirow{2}{*}{25,600} & \multirow{2}{*}{12,800} & \multirow{2}{*}{51,200} & \multirow{2}{*}{25,600}\tabularnewline
\multicolumn{2}{|c|}{($MBytes/s$)} &  &  &  &  &  & \tabularnewline
\hline 
\multirow{3}{*}{$N_{pu-OLS}$} & S5 & \multicolumn{2}{c|}{3-4} & \multicolumn{2}{c|}{1-2} & \multicolumn{2}{c|}{1}\tabularnewline
\cline{2-8}
 & A10 & \multicolumn{2}{c|}{3-4} & \multicolumn{2}{c|}{1-3} & \multicolumn{2}{c|}{1}\tabularnewline
\cline{2-8}
 & X+A10 & \multicolumn{2}{c|}{3-4} & \multicolumn{2}{c|}{1-2} & \multicolumn{2}{c|}{1}\tabularnewline
\hline 
\end{tabular}
\end{table*}

\begin{figure*}
\begin{centering}
\includegraphics[viewport=0bp 10bp 410bp 310bp,clip,scale=0.6]{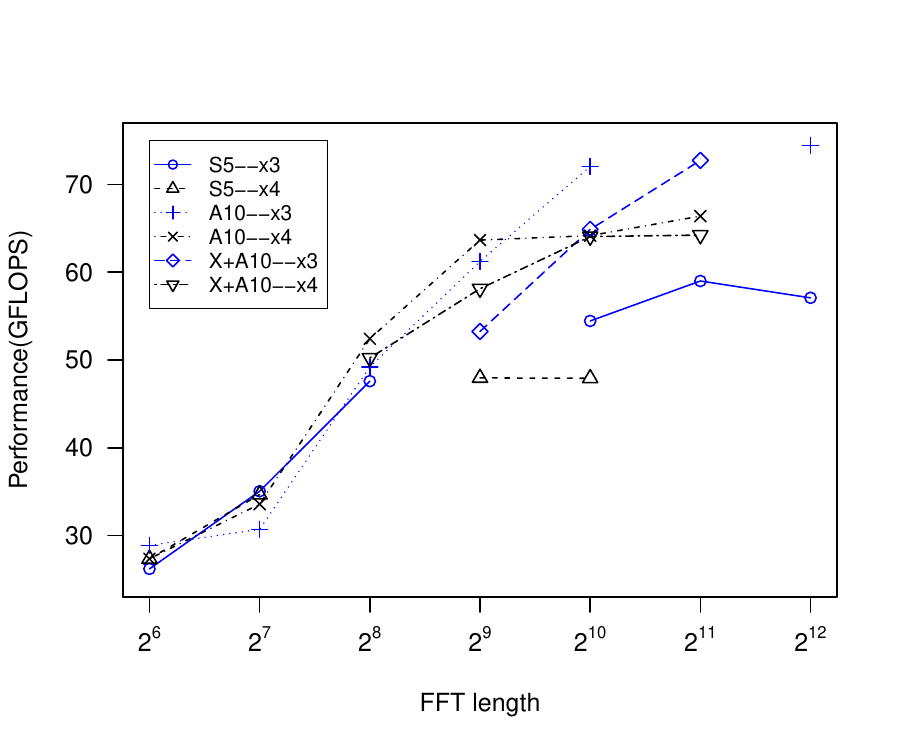}\includegraphics[viewport=0bp 10bp 410bp 310bp,clip,scale=0.6]{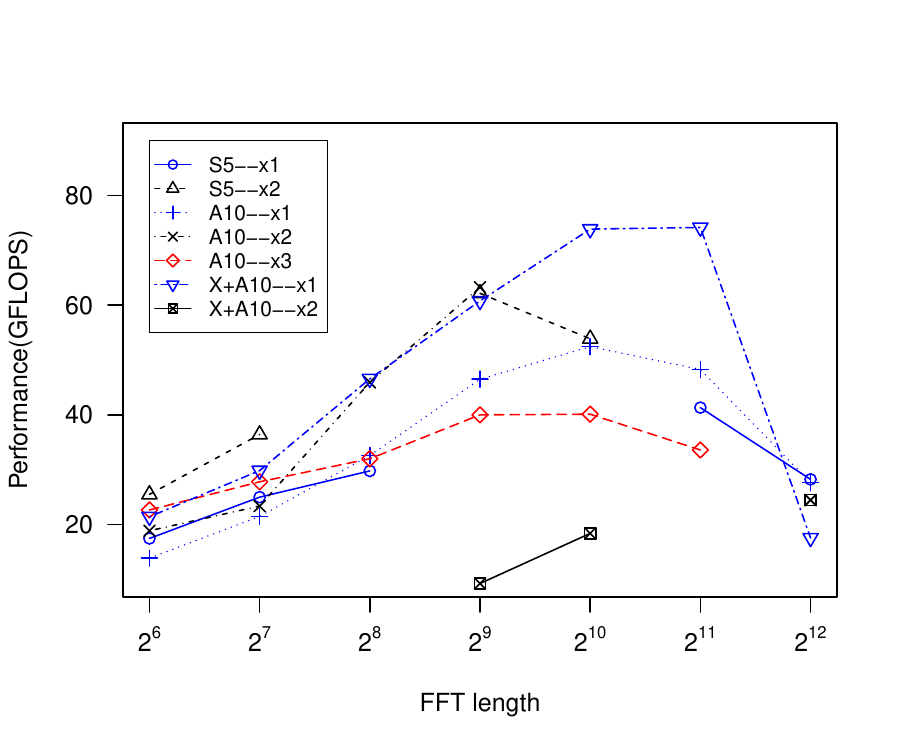}
\par\end{centering}
\caption{\label{fig:Performance4=0000268}Performance of multiple FD-OLS implementations
using 4-point (left) and 8-point (right) FFT engines}
\end{figure*}

\begin{figure}
\begin{centering}
\includegraphics[viewport=0bp 10bp 410bp 210bp,clip,scale=0.6]{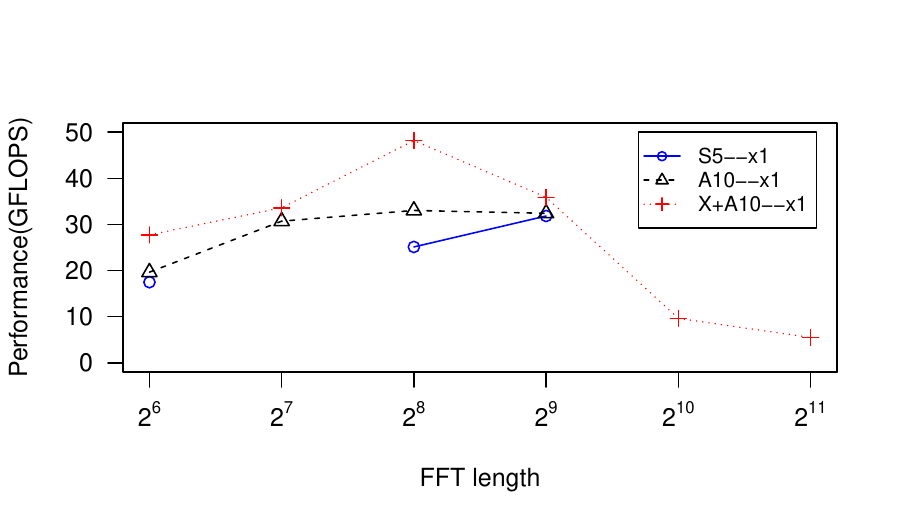}
\par\end{centering}
\caption{\label{fig:Performance-16p}Performance of multiple FD-OLS implementations
using 16-point FFT engine}
\end{figure}

With the $N_{pu-OLS}$ from Table~\ref{tab:Required-off-chip-memory},
the performance of the designs with different parameters are now evaluated.
The results are plotted in Figure~\ref{fig:Performance4=0000268}
(4-point and 8-point FFT engines) and Figure~\ref{fig:Performance-16p}
(16-point FFT engine). Not all of the designs can be compiled successfully,
in some cases even when the estimated resource usage is less than
100\%, but there was not clear pattern or other indicator from the
complier to why that happened. Although multiple FFT engines are implemented
in parallel, the performance in terms of $GFLOPS$ is about the same
as that of the performance of a single FFT engine. The main reason
is that a low occupancy percentage, which is less than 50\% for most
of the kernels. A low occupancy percentage implies that data cannot
be loaded or stored efficiently~\cite{intelopenclbest}, and it is
caused by a large number of streamed points per clock cycle required
by the kernel. 

Using the results of Figure~\ref{fig:Performance4=0000268} and Figure~\ref{fig:Performance-16p},
the FFT engine and $N_{pu-OLS}$ that provide the best performance
for a specific $N_{OLS-FT}$ are recorded in Table~\ref{tab:multiple_OLS_results}.
Following the analysis in Section~\ref{subsec:Theoretical-Analysis},
the larger the value of $N_{OLS-FT}$$/Tap_{N_{filter}}$, the more
operations can be saved. The optimal design for a specific matched
filter group might not be the design with the highest $GFLOPS$ in
Table~\ref{tab:multiple_OLS_results}. Taking S5 as an example, although
$N_{OLS-FT}=512$ based design provides a better performance than
other designs, $N_{OLS-FT}=2,048$ or $N_{OLS-FT}=4,096$ based designs
might be faster if $Tap_{N_{filter}}$ is close to 512. . Regarding
A10, the best performing design has the largest $N_{OLS-FT}$ among
the eventuated designs. In terms of X+A10, the $N_{OLS-FT}=2,048$
based design is over 2x times faster than $N_{OLS-FT}=4,096$ based
design, so it is the fastest among the evaluated designs.

\begin{table*}
\caption{\label{tab:multiple_OLS_results}Performance of the best designs}

\centering{}%
\begin{tabular}{|c|c|c|c|c|c|c|c|c|}
\hline 
Device & FD-OLS-($N_{OLS-FT}$) & \multicolumn{1}{c|}{64} & \multicolumn{1}{c|}{128} & \multicolumn{1}{c|}{256} & \multicolumn{1}{c|}{512} & \multicolumn{1}{c|}{1,024} & 2,048 & 4,096\tabularnewline
\hline 
\hline 
\multirow{4}{*}{S5} & $N_{pu-OLS}$ & 4 & 2 & 3 & 2  & 3 & 3 & 3\tabularnewline
\cline{2-9}
 & FFT engine  & 4-point & 8-point & 4-point & 8-point & 4-point & 4-point & 4-point\tabularnewline
\cline{2-9}
 & $f_{max}$$(MHz)$ & 217.6 & 216.7 & 217.8 & 218.2 & 219.7 & 204.3 & 186.9\tabularnewline
\cline{2-9}
 & Performance $(GFLOPS)$ & 27.35 & 36.4 & 47.6 & \textbf{62.2} & 54.5 & 59.0 & 57.1\tabularnewline
\hline 
\multirow{4}{*}{A10} & $N_{pu-OLS}$ & 3 & 4  & 4 & 2  & 3 & 4 & 3\tabularnewline
\cline{2-9}
 & FFT engine  & 4-point & 4-point & 4-point & 8-point & 4-point & 4-point & 4-point\tabularnewline
\cline{2-9}
 & $f_{max}$$(MHz)$ & 248.1 & 183.0 & 203.4 & 199.2 & 197.0 & 179.6 & 200.0\tabularnewline
\cline{2-9}
 & Performance $(GFLOPS)$ & 28.9 & 33.6 & 52.5 & 64.1 & 72.5 & 66.4 & \textbf{74.5}\tabularnewline
\hline 
\multirow{4}{*}{X+A10} & $N_{pu-OLS}$ & 1 & 1  & 3 & 1 & 1 & 4 & 2\tabularnewline
\cline{2-9}
 & FFT engine  & 16-point & 16-point & 4-point & 8-point & 8-point & 4-point & 8-point\tabularnewline
\cline{2-9}
 & $f_{max}$$(MHz)$ & 201.9 & 180.6 & 216.67 & 222.9 & 229.2 & 218.8 & 151.6\tabularnewline
\cline{2-9}
 & Performance $(GFLOPS)$ & 27.8 & 33.6 & 50.3 & 60.8 & 73.9 & \textbf{74.2} & 27.7\tabularnewline
\hline 
\end{tabular}
\end{table*}

\subsection{Comparison}

For the generic matched filter group implementation, the actual OpenCL
kernel is the same as that of the optimised design. The essential
difference is the number of launch times, which has been discussed
in Section~\ref{sec:Optimistion}. In this section, we therefore
focus on the TD vs. FD and FPGA vs. GPU comparisons.

\subsubsection{\label{subsec:TD-vs.-FD}TD vs. FD}

\begin{figure*}
\begin{centering}
\includegraphics[viewport=0bp 0bp 280bp 220bp,clip,scale=0.8]{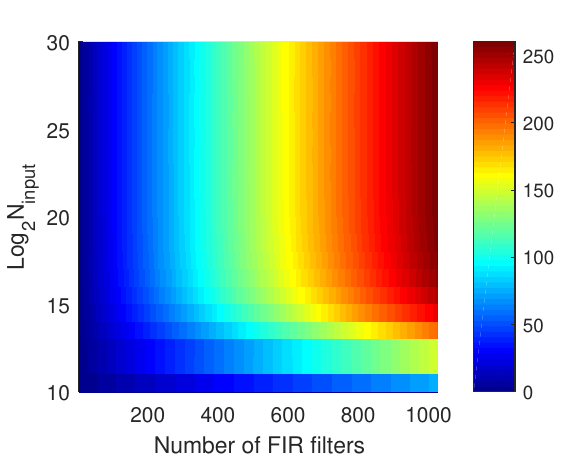}\includegraphics[viewport=0bp 0bp 280bp 220bp,clip,scale=0.8]{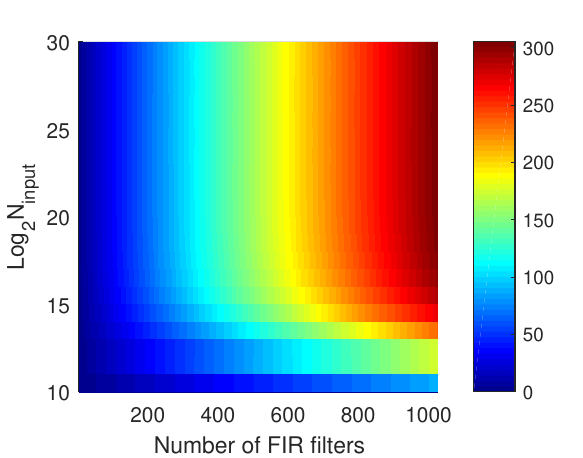}
\par\end{centering}
\caption{\label{fig:ols_vs_ola}Speedup of FD designs over TD designs on S5
(left) and A10 (right)}
\end{figure*}

Using the best $N_{OLA-tap}$ from Figure~\ref{fig:-real_Nopt} and
the evaluation results in Table~\ref{tab:multiple_OLS_results},
the TD and FD designs are compared in Figure~\ref{fig:ols_vs_ola},
depicting a colour map of the execution time speedup of FD over TD.
The designs on both S5 and A10 use $N_{OLS-FT}=4096$, and all $N_{pu-OLS}$
processing units share the same padded input. For most of the $MF$s,
the TD design performs worse than the FD design. The TD design only
beats the FD design for few $MF$s with $N_{input}=1024$. When $N_{filter}$
is over a hundred and $N_{input}$ is large than $2^{11}$, the FD
design is up to 250x times faster than the TD design on S5 and up
to 300x times faster on A10. The execution time of pre-processing,
such as Fourier transforming the input signals and coefficient arrays
are not considered.

\begin{figure}
\begin{centering}
\includegraphics[viewport=0bp 0bp 280bp 220bp,clip,scale=0.8]{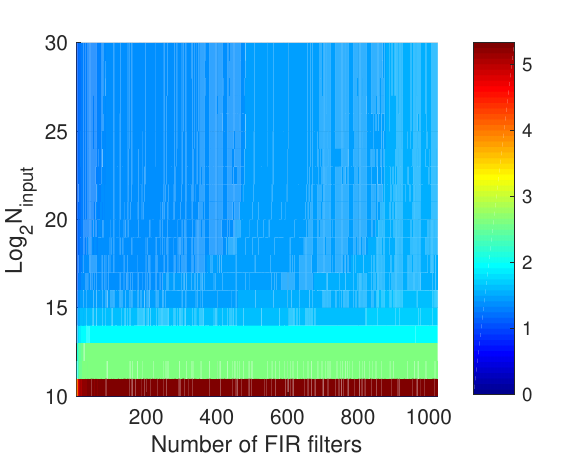}
\par\end{centering}
\caption{\label{fig:ols_ola_memory}Required off-chip memory of FD designs
relative to corresponding TD designs }
\end{figure}

The required off-chip memory is affected by the input size and the
number of filters, but also differs between the designs due to padding
For S5 and A10, the required memory of the FD designs relative to
that of the TD designs is shown in Figure~\ref{fig:ols_ola_memory}.
It can be seen that most FD designs need over 1.5x the off-chip memory
of TD designs. 

\subsubsection{FPGA vs. GPU}

\begin{figure}
\begin{centering}
\includegraphics[scale=0.65]{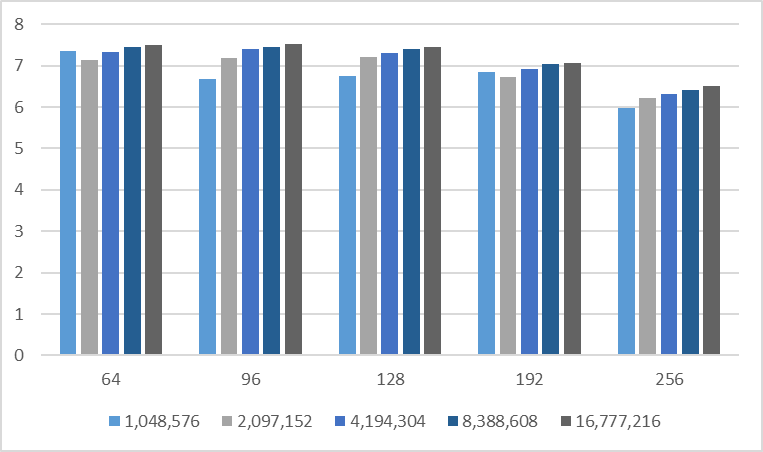}
\par\end{centering}
\caption{\label{fig:Speedup_gpu}Speedup of a single P100 card \cite{dimoudi2018gpu}
over a single A10 card}
\end{figure}

Based on the results in Section~\ref{subsec:TD-vs.-FD}, the proposed
FD design on A10 (FD-OLS with 4-point FFT engine, $N_{OLS-FT}=4,096$
and $N_{pu-OLS}=3$) is compared with optimised designs for high-end
GPUs, implementing the same matched filter groups. In~\cite{dimoudi2018gpu},
the latest NVIDIA Pascal GPU architecture with Tesla P100 card is
employed to evaluated a group of $MF$s, whose $N_{filter}$ ranges
from 64 to 256 and $N_{input}$ ranges from $2^{20}$ to $2^{24}$.
The power spectrum is calculated after Fourier transforming and a
custom GPU-based FFT engine, which is well-optimised for high-end
GPUs, was develop in~\cite{dimoudi2018gpu}. As can be seen in Figure~\ref{fig:Speedup_gpu},
a single P100 card performs up to 7.5x times faster than a single
A10 card. The performance lead drops to 6x-6.5x when the number of
filters in the group becomes quite large. The main advantages of P100
is that the supported memory bandwidth is over 10x times faster than
that of A10.

It is assumed that the TDP value is employed as GPU power consumption
due to the lack of other information. For most designs on A10, the
required power is less than $35W$~\cite{wangcombining}, which is
over 8.5x times lower than the TDP of P100, which is $300W$, hence
one can assume that the performance per/watt of the FPGA A10 and GPU
P100 is comparable. In terms of the price, P100 is 1.2x times more
expensive than that of A10 card, which is over 5,000 USD at the time
of writing..

\section{\label{sec:Conclusions}Conclusions}

This paper proposed optimised designs for multiple complex floating-point
filters of different sizes. We analytically studied matched filter
groups, where all filters have different sizes within a given range,
in both time-domain and Fourier-domain, using the overlap-add algorithm
and the overlap-save algorithm. Both TD-OLA and FD-OLS processing
can be significantly improved compared to a generic implementation
with a single (maximum) filter size, when optimising for the varying
size range. A load balancing algorithms was employed to process multiple
filters for TD-OLA designs and FD-OLS designs were significantly enhanced
carefully determining the best filter sizes and sub-grouping filters
to share the same padded input. The proposed designs were evaluated
on four types of FPGA devices. The best solution for a matched filter
group on a specific device was determined. When comparing frequency
domain with time domain designs, the multiple FD-OLS designs can be
over 250x times faster than the multiple TD-OLA designs, while using
1.5x times the off-chip memory. Lastly, we set the performance of
the FPGA design in relation to optimised designs for top-end GPUs.
While our proposed FPGA design on a mid-range FPGA was up to 7.5x
times slower than the well-optimised design on the top-end GPU, the
performance per watt was slightly in favour of our FPGA implementation.

\section*{Acknowledgment}

The authors acknowledge discussions with the TDT, a collaboration
between Manchester and Oxford Universities, and MPIfR Bonn and the
work benefited from their collaboration. We gratefully acknowledge
that this research was financially supported by the SKA funding of
the New Zealand government through the Ministry of Business, Innovation
and Employment (MBIE).

\bibliographystyle{plain}
\bibliography{mfg}

\end{document}